\documentclass[twocolumn,longbibliography,showpacs,superscriptaddress,nofootinbib,article]{revtex4-1}
\usepackage{graphicx}
\usepackage{bm}
\usepackage{amssymb}
\usepackage{color}
\usepackage{amsmath}
\usepackage{amstext}
\usepackage{latexsym}
\usepackage[usenames,dvipsnames]{xcolor}
\usepackage[colorlinks=true,citecolor=blue,linkcolor=RubineRed,urlcolor=blue]{hyperref}

\usepackage{phfqit}
\renewcommand{\DSym}{S}%

\usepackage{hyperref}%
\colorlet{linkcolor}{blue!50!black}
\hypersetup{bookmarksnumbered=false,bookmarksopen=false,bookmarksopenlevel=1,%
  breaklinks=true,pdfborder={0 0 0},colorlinks=true,%
  anchorcolor=linkcolor,citecolor=linkcolor,%
  filecolor=linkcolor,linkcolor=linkcolor,%
  menucolor=linkcolor,runcolor=linkcolor,%
  urlcolor=linkcolor}
\usepackage[all]{hypcap}%

\usepackage[nameinlink,capitalize]{cleveref}
\crefname{thm}{Theorem}{Theorems}

\begin{document}

\title{Dynamical dimerization phase in Jaynes-Cummings lattices}

\author{Rub\'en Pe\~na}
\affiliation{Departamento de F\'isica, Universidad de Santiago de Chile, 
Avenida Ecuador 3493, 9170124, Santiago, Chile}
\author{Felipe Torres}
\affiliation{Departamento de F\'isica, Facultad de Ciencias,
  Universidad de Chile, Casilla 653, Santiago, Chile 7800024}
 \affiliation{Center for the Development of Nanoscience and Nanotechnology, Estaci\'on Central, 9170124, Santiago, Chile}
\author{Guillermo Romero}
\email{guillermo.romero@usach.cl}
\affiliation{Departamento de F\'isica, Universidad de Santiago de Chile, 
Avenida Ecuador 3493, 9170124, Santiago, Chile}
\affiliation{Center for the Development of Nanoscience and Nanotechnology, Estaci\'on Central, 9170124, Santiago, Chile}

\date{\today}

\begin{abstract}
We report on an emergent dynamical phase of a strongly-correlated light-matter system, which is governed by dimerization processes due to short-range and long-range two-body interactions. The dynamical phase is characterized by the spontaneous symmetry breaking of the translational invariance and appears in an intermediate regime of light-matter interaction between the resonant and dispersive cases. We describe the quench dynamics from an initial state with integer filling factor of a finite-sized array of coupled resonators, each doped with a two-level system, in a closed and open scenario. The closed system dynamics has an effective Hilbert space description that allows us to demonstrate and characterize the emergent dynamical phase via time-averaged quantities, such as fluctuations in the number of polaritons per site and linear entropy. We prove that the dynamical phase is governed by intrinsic two-body interactions and the lattice topological structure. In the open system dynamics, we show evidence about the robustness of dynamical dimerization processes under loss mechanisms. Our findings can be used to determine the light-matter detuning range, where the dimerized phase emerges.\end{abstract}


\maketitle

\section{Introduction} 
The development of technology encompasses a broad range of opportunities
to harness quantum phenomena. For instance, it is possible to manipulate light-matter quasiparticles
or polaritons which have new properties such as stimulated scattering~\cite{PhysRevLett.81.3920,PhysRevB.62.R16263}, lasing~\cite{PhysRevB.66.085304,doi:10.1063/1.1494126,Kasprzak:2006aa}, parametric amplification~\cite{PhysRevLett.84.1547,PhysRevB.62.R4825,Saba:2001aa}, and superfluidity~\cite{PhysRevLett.93.166401,KAVOKIN2003187}. These characteristics can be used to enhance the experimental realization of polaritonic devices, such as semiconductor microcavities, where the coupling between quantum-well excitons and cavity photons gives rise to hybrid light-matter quasiparticles~\cite{Tsintzos:2008aa}. In the microwave regime, superconducting circuits based on Josephson junctions also allow to harness light-matter interaction for simulating strongly correlated phenomena with light~\cite{Nunnenkamp_2011,Houck:2012aa,PhysRevA.86.023837,PhysRevX.4.031043,PhysRevX.6.021044,PhysRevX.7.011016,PhysRevLett.121.043602,Ma:2019aa,PhysRevLett.122.183601,yanay2019realizing,Roushan1175}. The underlying physics of light-matter based quantum simulators is governed by the Jaynes-Cummings-Hubbard (JCH) model~\cite{PhysRevA.76.031805,Hartmann:2006aa,Greentree:2006aa} which describes the dynamics of coupled-resonator arrays (CRAs), each doped with a two-level system (TLS). In this case, the manipulation of polaritonic excitations locally depends on the detuning between the light and matter frequencies but also is largely influenced by the lattice structure. As the detuning increases from the resonant to the dispersive regime, the system transits from the Mott-insulating state characterized by the hybridization of light and matter states to a superfluid phase of photons~\cite{PhysRevA.76.031805,Hartmann:2006aa,Greentree:2006aa,PhysRevA.77.031803,PhysRevA.80.023811,PhysRevA.80.063838,Quach:11,PhysRevE.87.022104}. 

In this work, we demonstrate by using numerical calculation and an analytic model that during the phase transition from the Mott-insulating to superfluid state, a dynamical dimerization phase (DDP) emerges, which is characterized by the spontaneous symmetry breaking of the translational invariance. As dynamical dimerization, we refer to the dynamics of a finite-sized Jaynes-Cummings (JC) lattice that exhibits resonances related to the two-sites JC lattice. In order to identify the new regime of DDP, we analyze the purely unitary quench dynamics of few-body Jaynes-Cummings lattices. In particular, we consider a quantum quench from an initial state with integer filling factor in a Jaynes-Cummings dimer, which has been proven useful to simulating second-order like phase transitions from the Mott-insulating to superfluid phase~\cite{Figueroa2018}. Here, we introduce an effective Hilbert space in the two-excitations subspace using the criterion of discarding higher energy polaritonic states, which are out-of-resonance over the evolution~\cite{PhysRevA.80.023811,PhysRevA.91.043841}. This description allows us quantitative explanations for time-averaged quantities such as fluctuations in the number of polaritons per site and linear entropy. Besides, the computational cost is substantially diminished by using a reduced effective Hilbert space. As we extend the quench dynamics to complex finite-sized CRAs, we demonstrate the emergence of DDP, which is governed by intrinsic two-body interactions in the Jaynes-Cummings lattice. In the open system scenario, our numerical results show evidence about the robustness of dynamical dimerization processes under loss mechanisms, so our work may find inspiration for the observation of DDP within state-of-the-art quantum technologies such as superconducting circuits~\cite{PhysRevX.4.031043,PhysRevX.7.011016} and trapped ions~\cite{PhysRevA.80.060301,PhysRevLett.111.160501}. 

This paper is organized as follows. In Sec.~\ref{Model}, we introduce the JCH model and the polariton mapping. In Sec.~\ref{quench}, we describe the quench protocol for the closed JCH dimer. Here, we provide analytical expressions for time-averaged order parameters using an effective Hilbert space. In Sec.~\ref{DDP}, we highlight the emergence of a dynamical dimerization phase as we extend the one-dimensional Jaynes-Cummings lattice to three and four sites. Here, subsection~\ref{closed} describes DDP in a closed system, while in subsection~\ref{open}, we introduce loss mechanisms in the Jaynes-Cummings lattice and discuss their effects on the dynamical phase transition. Finally, in Sec.~\ref{conclusion}, we present our concluding remarks.

\section{The model}
\label{Model}
\begin{figure}[t]
\centering
\includegraphics[scale=0.28]{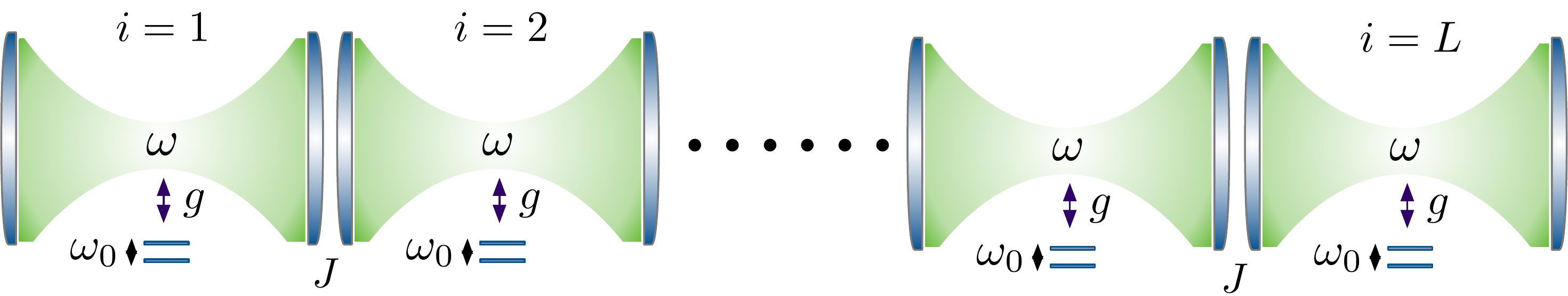}
\caption{Schematic representation of a Jaynes-Cummings-Hubbard lattice. Each resonator supports a single mode of the electromagnetic of frequency $\omega$, which interacts with a two-level system of frequency $\omega_0$. This interaction is represented by the coupling strength $g$. The interaction between cavities is characterized by the hopping parameter $J$.}
\label{Fig1}
\end{figure}

The Jaynes-Cummings-Hubbard model \cite{PhysRevA.76.031805,Hartmann:2006aa,Greentree:2006aa} describes a lattice of $L$ interacting coupled QED resonators, each supporting a single mode of the electromagnetic field which interacts with a two-level system. This situation is schematically shown in Fig.~\ref{Fig1}. The JCH Hamiltonian reads ($\hbar=1$)
\begin{align}
H_{\rm JCH} = \sum_iH^{\rm JC}_i-\sum_{\langle i,j\rangle}J_{ij}(a^{\dag}_ia_j+a^{\dag}_ja_i),
\label{HJCH}
\end{align}
where $a_i(a_i^{\dag})$ is the annihilation (creation) bosonic operator at the $i$th lattice site, $J_{ij}$ is the photon-photon hopping amplitude which takes values $J_{ij}=J$ if two sites of the lattice are connected and $J_{ij}=0$ otherwise. Also, $H^{\rm JC}_i=\omega a_i^{\dag}a_i + \omega_0\sigma_i^+\sigma_i^-+g(\sigma_i^+a_i+\sigma_i^-a_i^{\dag})$ is the Jaynes-Cummings (JC) Hamiltonian \cite{JC_Model} where $\sigma^{+}_i(\sigma^{-}_i)$ is the raising (lowering) operator acting on the $i$th TLS eigenbasis $\{\ket{\downarrow}_i,\ket{\uparrow}_i\}$, and $\omega$, $\omega_0$, and $g$ are the resonator frequency, TLS frequency, and light-matter coupling strength, respectively. 

In the resonant regime, $\Delta=\omega_0-\omega=0$, the hybridization of the light-matter yields to localized polariton excitations (Mott-insulator phase), while in the dispersive regime, $\Delta\gg g$, the system is dominated by photonic excitation behavior (superfluid phase)~\cite{PhysRevA.76.031805,Hartmann:2006aa}. This phase transition can also be described as a transition driven by the photon blockade effect from the Mott-insulator phase, where the intersite polariton exchange is forbidden, so effectively $J_{ij}=0$, to a superfluid phase dominated by a uniform photon hopping $J_{ij}=J$, in both cases, there is no cavity-embedded effect involved. As we demonstrate in Sec.~\ref{DDP}, the intermediate regime of light-matter interaction, which we define in the range $1<\Delta/g<10$, can be identified by the parameter $k_i=J(\sum_j\nu_{j})/L$ with $L$ the number of nonlinear coupled resonators, $J$ the hopping parameter, and $\nu_j$ the connectivity of node $j$. Notice that $k_i=0$ for the resonant case and $k_i=J$ for the dispersive case. This way, the origin of translational symmetry breaking can 
be explained by introducing a local order parameter of $\mu$-th phase 
$\Psi^{\mu}_i$, with $\mu=(\textrm{I},\textrm{II},\textrm{III})$ represents the resonant, intermediate, and dispersive case, respectively. Indeed, $\Psi^{\text{I}}_i(k_i)=\Psi^{\text{I}}(0)$, and $\Psi^{\text{III}}_i(k_i)=\Psi^{\text{III}}(J)$, which means that in the resonant and dispersive regimes there is translational invariance. Since the order parameter shows a spatial dependence in an intermediate regime 
$\Psi^{\text{II}}_i(k_i)$, then translational symmetry is broken. As a consequence, a \emph{dynamical dimerization phase} will happen due to intrinsic two-body interactions and the connectivity of each lattice site.

 The Hamiltonian (\ref{HJCH}) preserves the total number excitations (polaritons) described by the operator $\mathcal{N}=\sum_{i=1}^L(a^{\dag}_ia_i+\sigma^{+}_i\sigma^{-}_i)$. The $i$th node of the lattice in Fig.~\ref{Fig1} is described by the JC Hamiltonian $H^{\rm JC}_i$ whose eigenstates define the upper ($+$) and lower $(-)$ polaritonic basis $\ket{n,\pm}_i=\gamma_{n\pm}\ket{\downarrow,n}_i+\rho_{n\pm}\ket{\uparrow,n-1}_i$ with energies $E^{\pm}_n=n\omega+\Delta/2\pm\chi(n)$. Here,
$\chi(n)=\sqrt{\Delta^2/4+g^2n}$, $\rho_{n+}=\cos(\theta_n/2)$,
$\gamma_{n+}=\sin(\theta_n/2)$, $\rho_{n-}=-\gamma_{n+}$,
$\gamma_{n-}=\rho_{n+}$, $\tan\theta_n=2g\sqrt{n}/\Delta$, and the
detuning parameter $\Delta=\omega_0-\omega$. Also, one introduces the $i$th polaritonic
creation operators as $P^{\dag (n,\alpha)}_i=\ket{n,\alpha}_i\bra{0,-}$, where
$\alpha=\pm$ and we identify $\ket{0,-}\equiv\ket{\downarrow,0}$ and
$\ket{0,+}\equiv\ket{\emptyset}$ being a ket with all entries equal to
zero, that is, it represents an unphysical state. These
identifications imply $\gamma_{0-}=1$ and
$\gamma_{0+}=\rho_{0\pm}=0$. 

Using the above defined polaritonic basis, the
Hamiltonian (\ref{HJCH}) can be rewritten as
\cite{PhysRevA.76.031805,PhysRevA.80.023811}.
\begin{widetext}
\begin{eqnarray}
H = \sum^L_{i=1}\sum^{\infty}_{n=1}\sum_{\alpha=\pm}E_n^{\alpha} 
P^{\dag (n,\alpha)}_iP^{(n,\alpha)}_i
-\sum_{\langle i,j\rangle}J_{ij}\Big[\sum^{\infty}_{n,m=1}\sum_{\alpha,\alpha',\beta,\beta'=\pm}t_{n}^{\alpha\alpha'}
t_{m}^{\beta\beta'}P^{\dag (n-1,\alpha)}_iP^{(n,\alpha')}_iP^{\dag (m,\beta)}_jP^{(m-1,\beta')}_j+ {\rm H.c}\Big],\nonumber\\
\label{HPolariton}
\end{eqnarray}
\end{widetext}
where the matrix elements $t_{n}^{\alpha\alpha'}=\sqrt{n}\gamma_{(n-1)\alpha}\gamma_{n\alpha'}+\sqrt{n-1}\rho_{(n-1)\alpha}\rho_{n\alpha'}$. The first term in Eq.~(\ref{HPolariton}) stands for the local
polaritonic energy with an anharmonic spectrum and gives rise to an
effective on-site polaritonic repulsion. This is analog to the on-site photon repulsion in the Bose-Hubbard model~\cite{PhysRevB.40.546}. The last term in
Eq.~(\ref{HPolariton}) represents the polariton hopping between
resonators. This interaction also allows the interchange of polaritonic species of one or both sites involved~\cite{ PhysRevA.76.031805,PhysRevA.80.023811,PhysRevA.91.043841}, leading to a quite involve quantum dynamics.
  
A detailed understanding of the equilibrium properties of the JCH model (\ref{HPolariton}) resorts on approximated analytical solutions~\cite{PhysRevLett.103.086403} or numerical approaches such as density matrix renormalization group~\cite{PhysRevLett.99.186401,Rossini_2008,PhysRevA.88.063801,PhysRevB.96.174502}. In nonequilibrium situations, one can understand the underlying physics using the time-evolving block decimation algorithm~\cite{Grujic_2012,PhysRevLett.93.040502,PhysRevLett.93.207205}, or simplifying the description using effective Hilbert spaces~\cite{PhysRevA.78.063805,PhysRevA.91.043841,PhysRevA.80.043842,Chakrabarti_2011,Coto_2013,PhysRevA.88.053843}. The latter is particularly appropriate for studying the quench protocol presented in this article, as we consider the closed system scenario. 

\section{Quench dynamics in a Jaynes-Cummings-Hubbard dimer}
\label{quench}
In this section, we introduce a sudden quench protocol~\cite{Figueroa2018} and its effects on the quantum dynamics of the JCH dimer. Also, we provide quantitative explanations of time-averaged quantities such as fluctuations in the number of polaritons per site and linear entropy using an effective Hilbert space. In Sec.~\ref{DDP}, we will show that the underlying physics of the quench dynamics allows us to understand the emergent \emph{dynamical dimerization phase}. We stress that DDP occurring in the JCH lattice happens in the frequency regime $Jn\ll g\sqrt{n}\ll \omega n$~\cite{PhysRevA.76.031805}, where the rotating wave approximation holds.
\begin{figure}[t]
\centering
\includegraphics[scale=0.39]{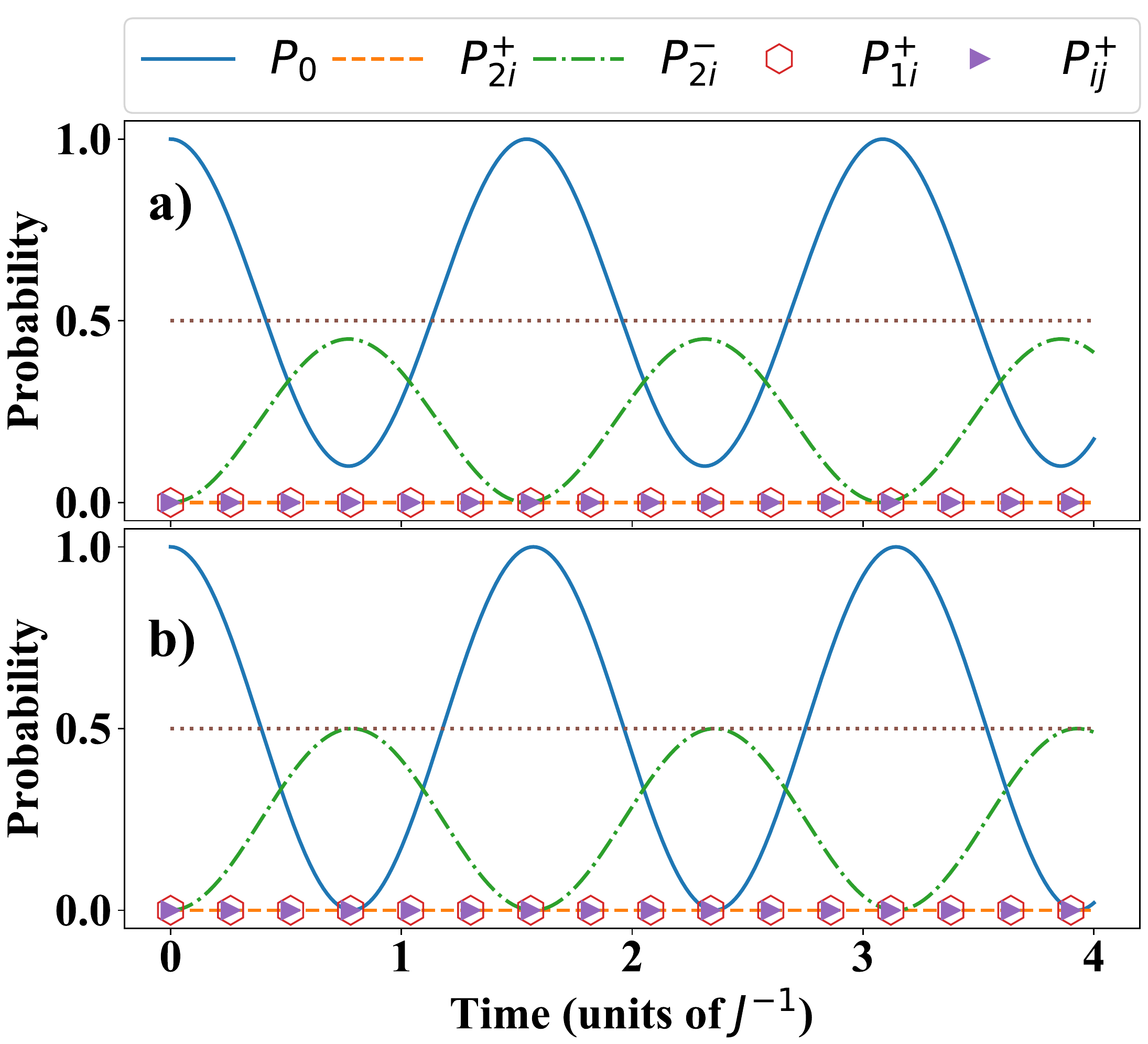}
\caption{Populations of states $\ket{\psi_0}$, $\ket{\psi^{\pm}_2}_i$, $\ket{\psi^{+}_1}_i$, and $\ket{\psi^{+}_1}_{ij}$, defined in the main text, as a function of time. The dimer is initialized in the state $\ket{\psi_0}$, parameters are $g=10^{-2}\omega$, $J=10^{-4}\omega$, where $\omega$ is the resonator frequency, and we consider up to $5$ Fock states per resonator. (a) $\Delta=5g$, (b) $\Delta=50g$. Horizontal dotted lines have been added as a guide to the eye.}
\label{Fig2}
\end{figure} 

For each detuning $\Delta$, we set the initial condition as the lowest energy state with integer filling factor of one excitation per site, that is, $\ket{\psi_0}=\otimes^{L}_{i=1}\ket{1,-}_i$ ($L$ is the number of lattice sites) which corresponds to a Mott-insulating state at hopping rate $J=0$. Then, at time $t=0$, the parameter $J$ is suddenly quench to a new value $J_f\ne 0$ such that the Hamiltonian has changed to $H=H_{\rm JCH}(J_f)$. Hence, the JCH lattice dynamics is described by the state $\ket{\psi_0(t)}=e^{-iHt}\ket{\psi_0}$ $(\hbar=1)$ which leads to nonequilibrium phenomena. In order to characterize DDP, we compute time-averaged order parameters such as the variance of the number of polaritons per site ${\rm Var}(n_i)=\frac{1}{\tau}\int_{0}^{\tau}dt(\langle n^2_i\rangle-\langle n_i\rangle^2)$, where $n_i=a^{\dag}_ia_i+\sigma^{+}_i\sigma^{-}_i$, and $\tau=J^{-1}$, or the linear entropy $E=\frac{1}{\tau}\int_{0}^{\tau}dtS_{\rho_{i}}(t)$, where $S_{\rho_i}(t)=1-{\rm Tr}(\rho^2_i)$, and $\rho_i$ the reduced density matrix of the leftmost or rightmost site of the JC dimer.

The quench protocol described above allows to simulating second-order like phase transition captured via the ${\rm Var}(n_i)$, see Ref.~\cite{Figueroa2018}, which is analog to the adiabatic dynamics studied in Ref.~\cite{PhysRevA.76.031805}. Also, the simulated phase transitions in a JCH lattice can be characterized via linear entropy, as shown in this article, which implies that the observation of the emergent DDP is independent of the choice of the order parameter.

\emph{Effective Hilbert space for a closed system.}\textemdash In order to introduce an effective description of the system dynamics, it is useful to consider the JCH Hamiltonian written in the polaritonic basis, see Eq.~(\ref{HPolariton}). Starting from the inital state $\ket{\psi_0}=\ket{1,-}_i\ket{1,-}_j$, the JCH Hamiltonian (\ref{HPolariton}) may lead to processes such as the exchange of polaritonic species, or the interchange of polaritonic species of one or both sites involved ($i,j$). In this case, the full quantum dynamics should involve all states within the two-excitations subspace, which we define as $\{\ket{\psi_0}\!\!=\!\!\ket{1,-}_i\ket{1,-}_j,\ket{\psi^{\pm}_2}_i\!\!=\!\!\ket{2,\pm}_i\ket{0,-}_j,\ket{\psi^{\pm}_2}_j\!\!=\!\!\ket{0,-}_i\ket{2,\pm}_j,\ket{\psi^+_1}_i\!\!=\!\!\ket{1,+}_i\ket{1,-}_j,\ket{\psi^+_1}_j\!\!=\!\!\ket{1,-}_i\ket{1,+}_j,\ket{\psi^+_1}_{ij}\!\!=\!\!\ket{1,+}_i\ket{1,+}_j\}$. However, interchange of polaritonic excitations can be neglected under the conditions $\{|E^+_{2}-2E^-_1|,|2E^+_{1}-E^-_2|,|E^+_{1}+E^-_{1}-E^{-}_2|\}\gg J$, which results in fast oscillating contributions, and we can apply the rotating-wave approximation~\cite{PhysRevA.76.031805,PhysRevA.91.043841,PhysRevA.80.023811}. Figure~\ref{Fig2} shows that the interchange of polaritonic species is suppressed over the evolution. Here we plot the populations of above defined states as a function of time. We identify populations as $P_0=|\braket{\psi_0}{\psi(t)}|^2$, $P^{\pm}_{2i}=|_i\braket{\psi^{\pm}_{2}}{\psi(t)}|^2$, $P^{+}_{1i}=|_i\braket{\psi^{+}_{1}}{\psi(t)}|^2$, and $P^{+}_{ij}=|_{ij}\braket{\psi^{+}_{1}}{\psi(t)}|^2$. Due to the symmetry of the JC dimer, it is cleat that $P^{\pm}_{2j}=|_j\braket{\psi^{\pm}_{2}}{\psi(t)}|^2=P^{\pm}_{2i}$ and $P^{+}_{1j}=|_j\braket{\psi^{+}_{1}}{\psi(t)}|^2=P^{+}_{1i}$ (not shown in Fig.~\ref{Fig2}). In this work, we carry out numerical simulations with the quantum toolbox in Python QuTiP~\cite{Qutip,Qutip2}.

Since the interchange of polaritonic species is suppressed over the evolution, we can introduce an effective Hilbert space involving states of the lower polaritonic branch $\mathcal{H}_I=\{\ket{\psi_0},\ket{\psi^-_2}_i,\ket{\psi^-_2}_j\}$ for describing the quench dynamics. In this case, the effective Hamiltonian reads ($\hbar=1$)
\begin{equation}
H_{\rm eff}=
\left(
\begin{array}{ccc}
a& b & b\\
b & c & 0 \\
b & 0 & c \\
\end{array}
\right),
\label{Effective1}
\end{equation}
where $a=2E_{1}^-$, $b=-Jt_{1}^{--}t_{2}^{--}$, $c=E_{2}^-$, and $t_{1}^{--}t_{2}^{--}=\cos(\theta_1/2)(\sqrt{2}\cos(\theta_1/2)\cos(\theta_2/2)+\sin(\theta_1/2)\sin(\theta_2/2))$. Thus, the full dynamics can be solved analytically by diagonalizing the above $3\times 3$ matrix.
\begin{figure}[t]
\centering
\includegraphics[scale=0.42]{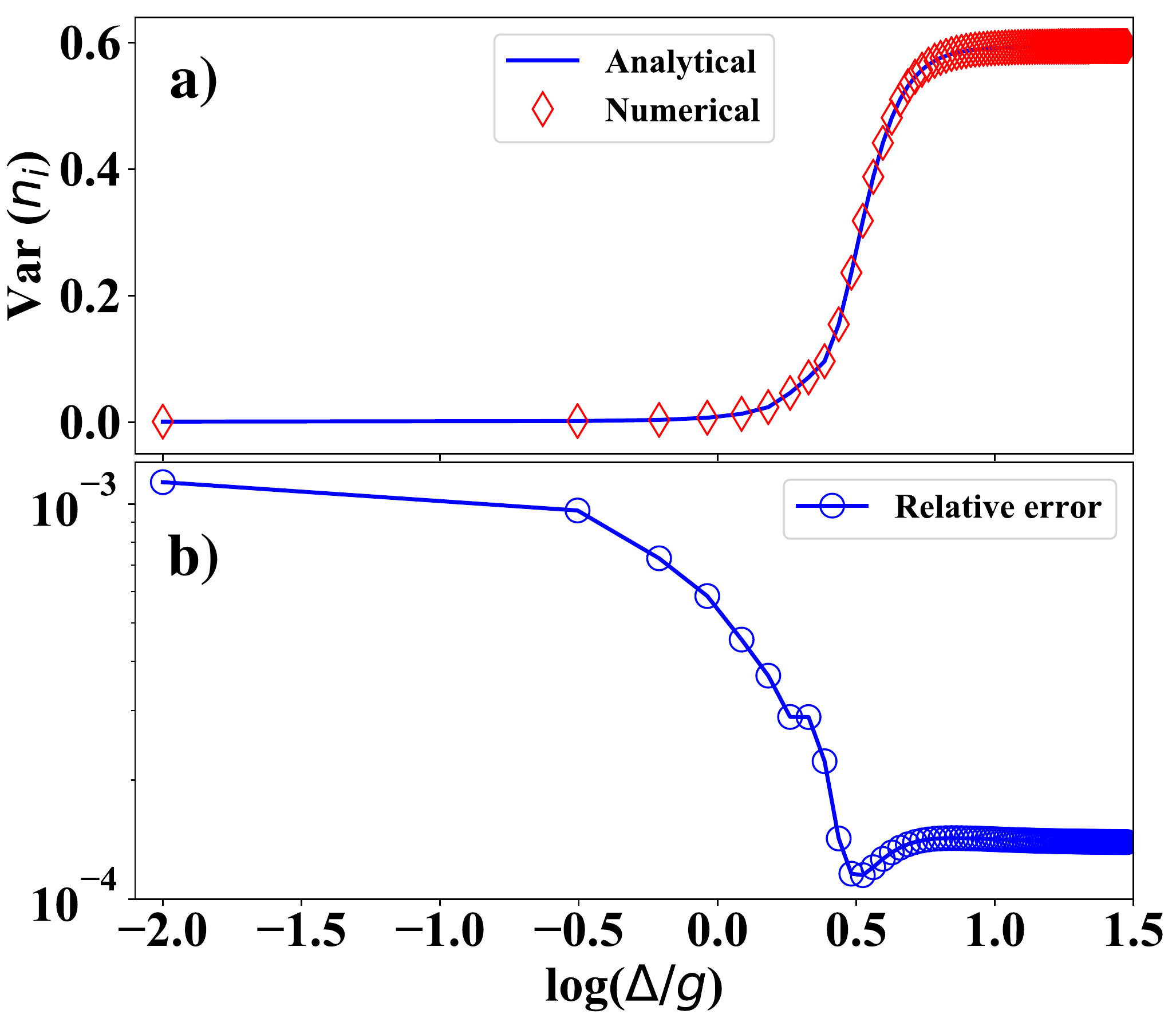}
\caption{(a) Time-averaged order parameter ${\rm Var}(n_i)$ as a function of ${\rm log}_{10}(\Delta/g)$. (b) Comparison between the analytical and numerical predictions of ${\rm Var}(n_i)$ (cf. Eq.~(\ref{varianceI})). We use parameters $g=10^{-2}\omega$, $J=10^{-4}\omega$, where $\omega$ is the resonator frequency.}
\label{Fig3}
\end{figure} 

Starting from the initial condition $\ket{\psi_0}=\ket{1,-}_i\ket{1,-}_j$, the wave function at time $t$ reads 
\begin{align}
\ket{\psi(t)}=c_{0}(t)\ket{\psi_0}+c_{2i}^-(t)\ket{\psi_2^-}_i+c_{2j}^-(t)\ket{\psi_2^-}_j,
\label{KetT}
\end{align}  
where the probability amplitudes are
\begin{subequations}
\begin{align}
c_0(t)&=\dfrac{1}{\alpha_+-\alpha_-}(\alpha_+e^{-it\lambda_+}-\alpha_-e^{-it\lambda_-}),\\
c_{2i}^-(t)&=c_{2j}^-(t)=\dfrac{1}{\alpha_+-\alpha_-}(e^{-it\lambda_+}-e^{-it\lambda_-}),
\end{align}
\label{amplitudes}
\end{subequations}
and we define the coefficients $\lambda_{\pm}\!\!=\!\!(a+c\pm\sqrt{8b^2+(a-c)^2})/2$, $\alpha_{\pm}\!=(\!a-c\pm\sqrt{8b^2 + (a-c)^2})/2b$. 

\emph{Time-averaged order parameters.}\textemdash Given the wave function 	(\ref{KetT}), we can analytically compute time-averaged order parameters such as the variance of the number of polaritons per site ${\rm Var}(n_i)=\frac{1}{\tau}\int_{0}^{\tau}(\langle n^2_i\rangle-\langle n_i\rangle^2)dt$, where $n_i=a^{\dag}_ia_i+\sigma^{+}_i\sigma^{-}_i$, or the linear entropy $E=\frac{1}{\tau}\int_{0}^{\tau}S_{\rho_{i}}(t)dt$, where $S_{\rho_i}(t)=1-{\rm Tr}(\rho^2_i)$, and $\rho_i$ the reduced density matrix of the leftmost or rightmost site of the JC dimer. Thus, the time-averaged variance reads 
\begin{align}
{\rm Var}(n_i)=\frac{4b^2}{\Omega_0^2}\Bigg[1-\frac{J}{\Omega_0}\sin\bigg(\frac{\Omega_0}{J}\bigg)\Bigg],
\label{varianceI}
\end{align}
where we define $\Omega_0=\sqrt{8b^2+(a-c)^2}$. 
\begin{figure}[t]
\centering
\includegraphics[scale=0.42]{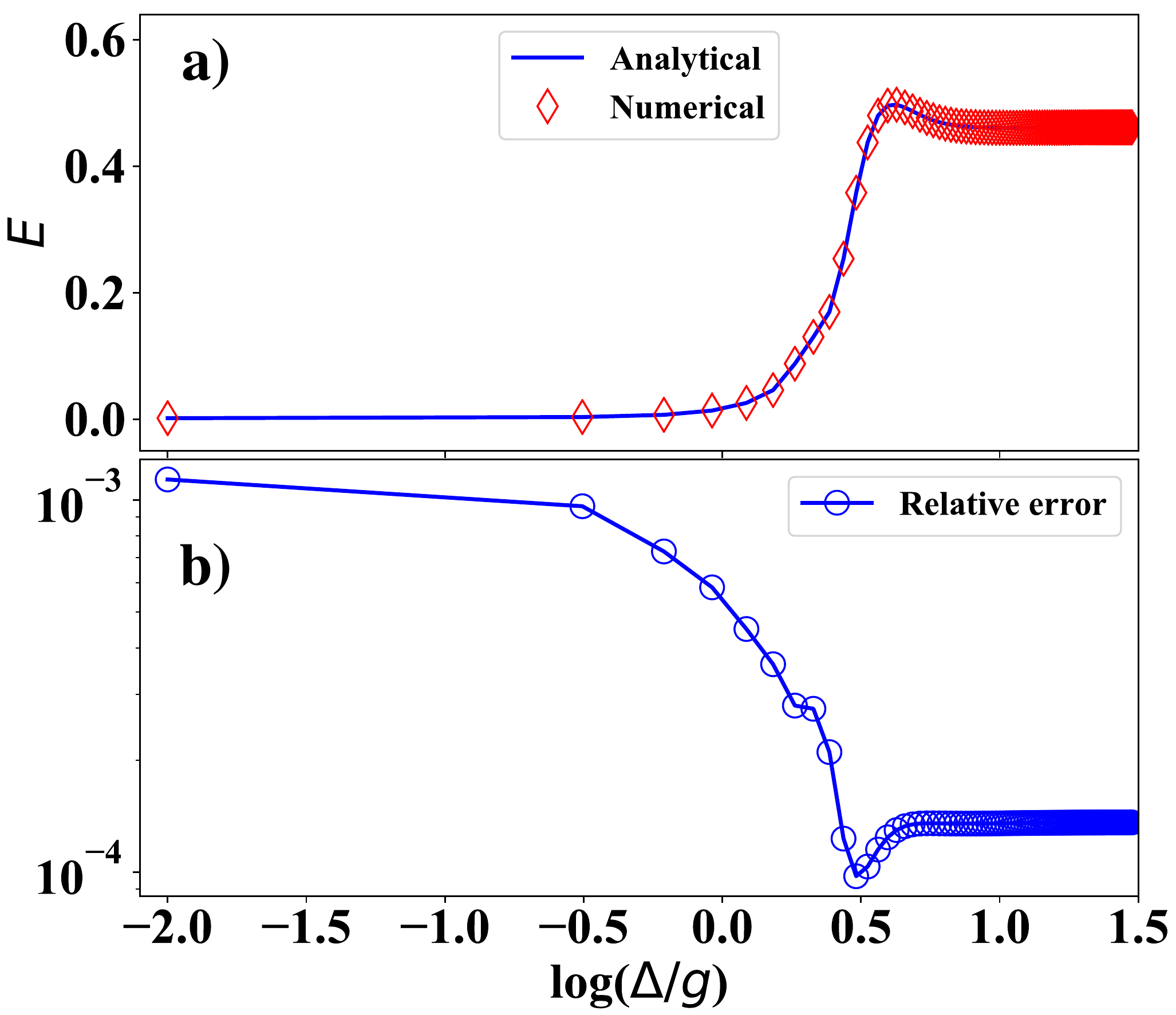}
\caption{(a) Time-averaged linear entropy $E$ as a function of ${\rm log}_{10}(\Delta/g)$. (b) Comparison between the analytical and numerical predictions of $E$ (cf. Eq.~(\ref{entropy})). We use parameters $g=10^{-2}\omega$, $J=10^{-4}\omega$, where $\omega$ is the resonator frequency.}
\label{Fig4}
\end{figure} 

Figure \ref{Fig3}(a) shows the behavior of ${\rm Var}(n_i)$ as a function of ${\rm log}_{10}(\Delta/g)$ calculated from the full numerics (red diamonds) and the analytical prediction (continuos blue line) in Eq.~(\ref{varianceI}). We see a good agreement between both predictions as the relative error shows in Fig.~\ref{Fig3}(b). Also, Eq.~(\ref{varianceI}) allows us to predict the asymptotic behavior of ${\rm Var}(n_i)$ as the detuning increases, $\Delta/g\to\infty$. In this case, the spectrum of the lower (upper) polaritonic branch becomes harmonic with eigenenergies $E_n^-\approx n\omega_R$ ($E_n^+\approx n\omega_R+\Delta$), where $\omega_R=\omega-g^2/\Delta$, thus allowing the resonance condition $E^-_2-2E^-_1=0$ ($a=c$) (cf. Fig.\ref{Fig2}(b)). Also, $|b|=\sqrt{2}J$ and $\Omega=4J$, so the asymptotic value of ${\rm Var}(n_i)$ reads
\begin{align} 
\lim_{\Delta/g\to\infty} {\rm Var}(n_i)=\frac{1}{2}\Bigg(1-\frac{1}{4}\sin4\Bigg)=0.5946.
\end{align}

It is worth mentioning that the analytical result (\ref{varianceI}) represents the hallmark for the dimer dynamics. In Sec.~\ref{DDP}, we will prove that as one increases the number of lattice sites, the time-averaged variance (\ref{varianceI}) allows us to identify resonances due to intrinsic short- and long-range two-body interactions, which govern the DDP.  

On the other hand, one can also characterize the dimer dynamics via the linear entropy of the reduced density matrix as mixedness measure~\cite{Giuliano-Rossini_Book}. First, notice that the quantum state (\ref{KetT}) is already in its Schmidt decomposition, which leads to a diagonal reduced density matrix 
\begin{equation}
\rho_i=\left(
\begin{array}{ccc}
|c^-_{2j}(t)|^2&0&0 \\
0&|c_{0}(t)|^2&0 \\
0&0&|c^-_{2i}(t)|^2
\end{array}
\right).
\end{equation}
In this case, the linear entropy as a function of time is given by $S_{\rho_i}(t)=1-(|c_0(t)|^4+2|c^-_{2i}(t)|^4)$, and the time-averaged linear entropy reads
\begin{align}
E=\frac{2b^2}{\Omega_0^5}\Bigg[2\Omega_0\Omega^2_1-4J\Omega^2_2\sin\bigg(\frac{\Omega_0}{J}\bigg)-3b^2J\sin\bigg(\frac{2\Omega_0}{J}\bigg)\Bigg],\nonumber\\
\label{entropy}
\end{align}
where $\Omega_1\!=\!\sqrt{7b^2+2(a-c)^2}$ and $\Omega_2\!=\!\sqrt{2b^2+(a-c)^2}$. Figure \ref{Fig4}(a) shows the behavior of $E$ as a function of ${\rm log}_{10}(\Delta/g)$ calculated from the full numerics (red diamonds) and the analytical prediction (continuos blue line) in Eq.~(\ref{entropy}). We see a good agreement between analytical and numerical predictions as shown in Fig.~\ref{Fig4}(b). We can also estimate the asymptotic value of the time-averaged entropy as one increases the ratio $\Delta/g$, it reads 
\begin{align} 
\lim_{\Delta/g\to\infty} E=0.4616.
\end{align}

In what follows, we will use the time-averaged variance (6) for demonstrating the existence of DDP in the intermediate regime of light-matter interaction. This choice establishes the physical framework for the subsequent discussion, but a similar analysis with the linear entropy leads to the same conclusion about DDP. 

\section{Dynamical dimerization phase}
\label{DDP}
\subsection{Closed system}
\label{closed}

Emergent dynamical critical phenomena following a quantum quench have experienced much interest in recent years~\cite{Heyl_2018,Heyl_2019,PhysRevLett.110.135704,PhysRevLett.120.130601,Halimeh2,Halimeh3,Halimeh4,PhysRevB.97.174401,Halimeh2018,PhysRevLett.120.130601,moeckel2008interaction,moeckel2010crossover,sciolla2011dynamical,gambassi2011quantum,sciolla2013quantum,maraga2015aging,mori2018thermalization,Zhang:2017ab,Halimeh1,PhysRevLett.110.135704,karrasch2013dynamical,andraschko2014dynamical,PhysRevLett.110.135704,jurcevic2017direct,PIROLI2018454,PhysRevLett.121.130603}. Here, we demonstrate an emergent \emph{dynamical dimerization phase} as we extend the Jaynes-Cummings  lattice to three (trimer) and four (tetramer) sites (cf.~Fig.\ref{Fig1}), and in an intermediate regime of light-matter coupling strength, that is, $1<\Delta/g<10$. We define the concept of dimerization as the dynamical process where short-range and long-range two-body interactions govern the quantum dynamics. Here, short-range (long-range) two-body interaction is due to the direct (mediated) exchange of polaritons. We will prove that the connectivity associated with each node of the lattice plays a crucial role to define DDP. 

Let us start the discussion with a three-sites JC lattice initialized in the state $\ket{\psi_0}=\ket{1,-}_i\ket{1,-}_j\ket{1,-}_k$, where the subindexes $i$, $j$, and $k$ refer to the leftmost, center, and rigthmost lattice site, respectively. If we let the system evolves according to the quantum quench described is Sec.~\ref{quench}, one should impose the conditions for neglecting interchange of polaritonic species between nearest-neighbor sites, that is, $\{|E^+_{2}-2E^-_1|,|2E^+_{1}-E^-_2|,|E^+_{1}+E^-_{1}-E^{-}_2|,|E^+_{1}-E^{-}_1|,|E^+_{3}-E^{-}_2-E^{-}_1|,|E^+_{2}-E^{-}_2|,|E^+_{2}+E^{+}_1-E^{-}_2-E^{-}_1|\}\gg J$. As for the dimer, only the lower polaritonic branch will be activated and the dimension of the effective Hilbert space ($\mathcal{H}$) is given by $(N+d-1)!/N!(d-1)!$, where $N$ is the number of excitations that should be distributed into $d$ lattice sites. In our case, $N=3$ and $d=3$ results in ${\rm dim}(\mathcal{H})=10$. At time $t$, the wave function may be written as a linear combination of states belonging to the three-excitations subspace, that is,
\begin{align}
\ket{\psi(t)}=&c_0(t)\ket{\psi_0}+c_{3j}\ket{\psi_{3j}}\nonumber\\
+&c_{3i}(t)(\ket{\psi_{3i}}+\ket{\psi_{3k}})\nonumber\\
+&c_{2i1j}(t)(\ket{\psi_{2i1j}}+\ket{\psi_{1j2k}})\nonumber\\
+&c_{1i2j}(t)(\ket{\psi_{1i2j}}+\ket{\psi_{2j1k}})\nonumber\\
+&c_{2i1k}(t)(\ket{\psi_{2i1k}}+\ket{\psi_{1i2k}}),
\label{trimerstate}
\end{align}
where we define states $\ket{\psi_0}=\ket{1,-}_i\ket{1,-}_j\ket{1,-}_k$, $\ket{\psi_{3i}}=\ket{3,-}_i\ket{0,-}_j\ket{0,-}_k$, $\ket{\psi_{3j}}=\ket{0,-}_i\ket{3,-}_j\ket{0,-}_k$, $\ket{\psi_{3k}}=\ket{0,-}_i\ket{0,-}_j\ket{3,-}_k$, $\ket{\psi_{2i1j}}=\ket{2,-}_i\ket{1,-}_j\ket{0,-}_k$, $\ket{\psi_{1j2k}}=\ket{0,-}_i\ket{1,-}_j\ket{2,-}_k$, $\ket{\psi_{1i2j}}=\ket{1,-}_i\ket{2,-}_j\ket{0,-}_k$, $\ket{\psi_{2j1jk}}=\ket{0,-}_i\ket{2,-}_j\ket{1,-}_k$, $\ket{\psi_{2i1k}}=\ket{2,-}_i\ket{0,-}_j\ket{1,-}_k$, and $\ket{\psi_{1i2k}}=\ket{1,-}_i\ket{0,-}_j\ket{2,-}_k$. Notice that some probability amplitudes are equal due to symmetry of the trimer with respect to the lattice center $(j)$. Here, the state (\ref{trimerstate}) will be computed via full numerics.  
\begin{figure}[t]
\centering
\includegraphics[scale=0.22]{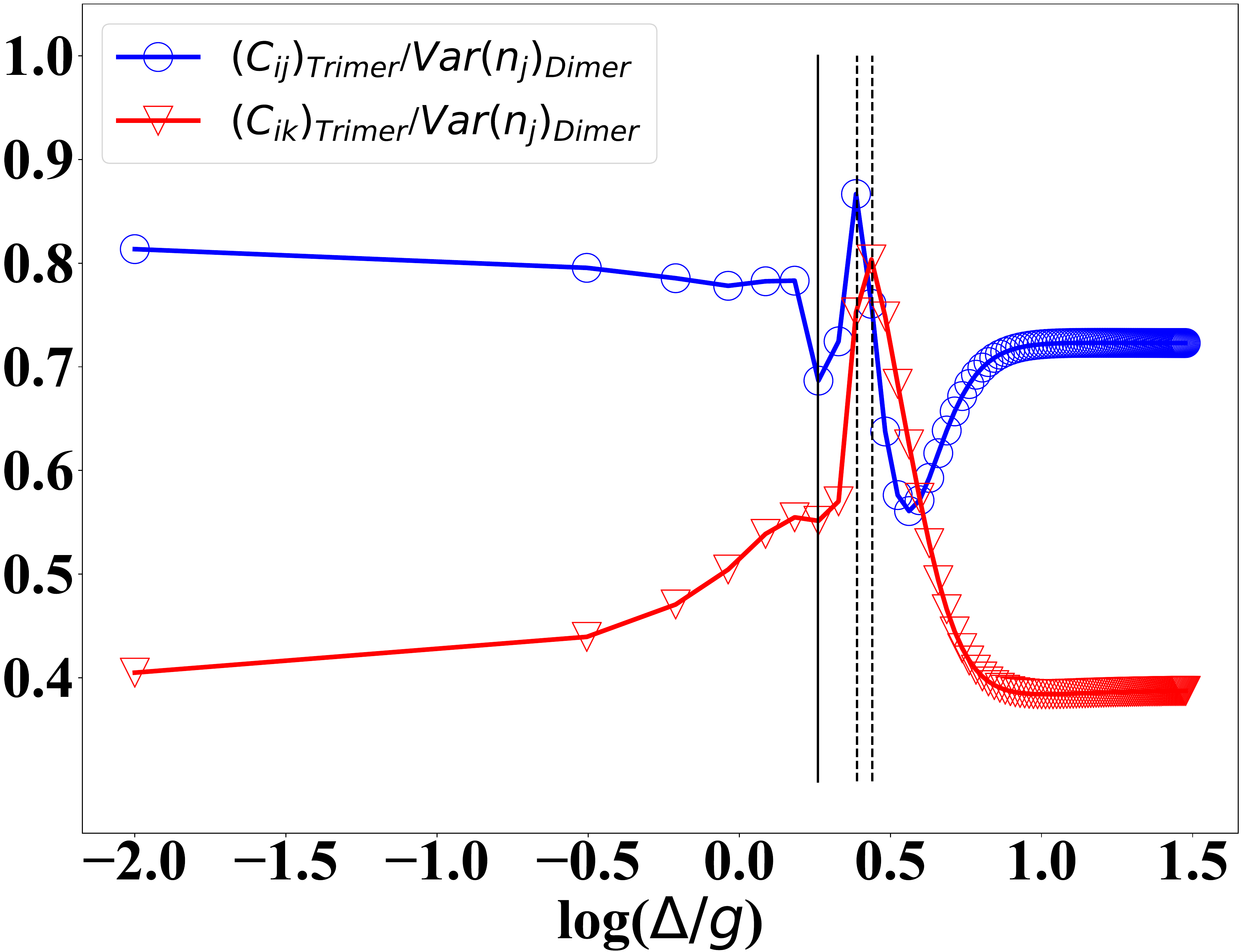}
\caption{Ratio between the absolute value of time-averaged nearest-neighbor correlation function $C_{ij}$ (next nearest-neighbor correlation function $C_{ik}$) of the trimer, and the dimer variance in Eq.~(\ref{varianceI}). Vertical dashed lines, from left to right, indicate critical values of detuning $\Delta/g=(2.43, 2.73)$ where dynamical dimerization processes happen. Continuous vertical line stands for the anti-resonance. We use parameters $g=10^{-2}\omega$, $J=10^{-4}\omega$, where $\omega$ is the resonator frequency, and we consider up to $5$ Fock states per resonator.}
\label{Fig5}
\end{figure}

Figure~\ref{Fig5} shows the ratio between the absolute value of the time-averaged nearest-neighbor correlation function $C_{ij}$ (next nearest-neighbor correlation function $C_{ik}$) of the trimer, and the dimer variance~(\ref{varianceI}). Two-point correlation functions are $C_{ij(k)}=\frac{1}{\tau}\int_{0}^{\tau}dt(\langle n_in_{j(k)}\rangle-\langle n_i\rangle \langle n_{j(k)}\rangle)$, where $\tau=J^{-1}$. Here, we identify two critical values of detuning, vertical dashed lines, $\Delta/g=(2.43, 2.73)$ within the intermediate regime of light-matter interaction, $1<\Delta/g<10$. At these critical points the trimer experiences dynamical dimerization processes, where short-range correlations rule the dynamics at $\Delta/g=2.43$, while at $\Delta/g=2.73$ a combination of both short- and long-range correlations govern the dynamics. The resonances shown in Fig.~\ref{Fig5} demonstrate that the intrinsic dimer dynamics, characterized by the time-averaged variance (\ref{varianceI}), governs the quench dynamics of the trimer. Furthermore, the dimer variance allows to identify short-range and long-range two-body interactions. The former is a consequence of direct cavity-cavity coupling of sites $(i,k)$, while the latter results from an indirect interaction between sites $(i,k)$ mediated by the center lattice site $j$. Notice that Fig.~\ref{Fig5} also exhibits an anti-resonance at $\Delta=1.82g$ (continuos vertical line). At this point, no dynamical dimerization happens, and the JC lattice remains approximately in the Mott insulating state.

\begin{figure}[t]
\centering
\includegraphics[scale=0.22]{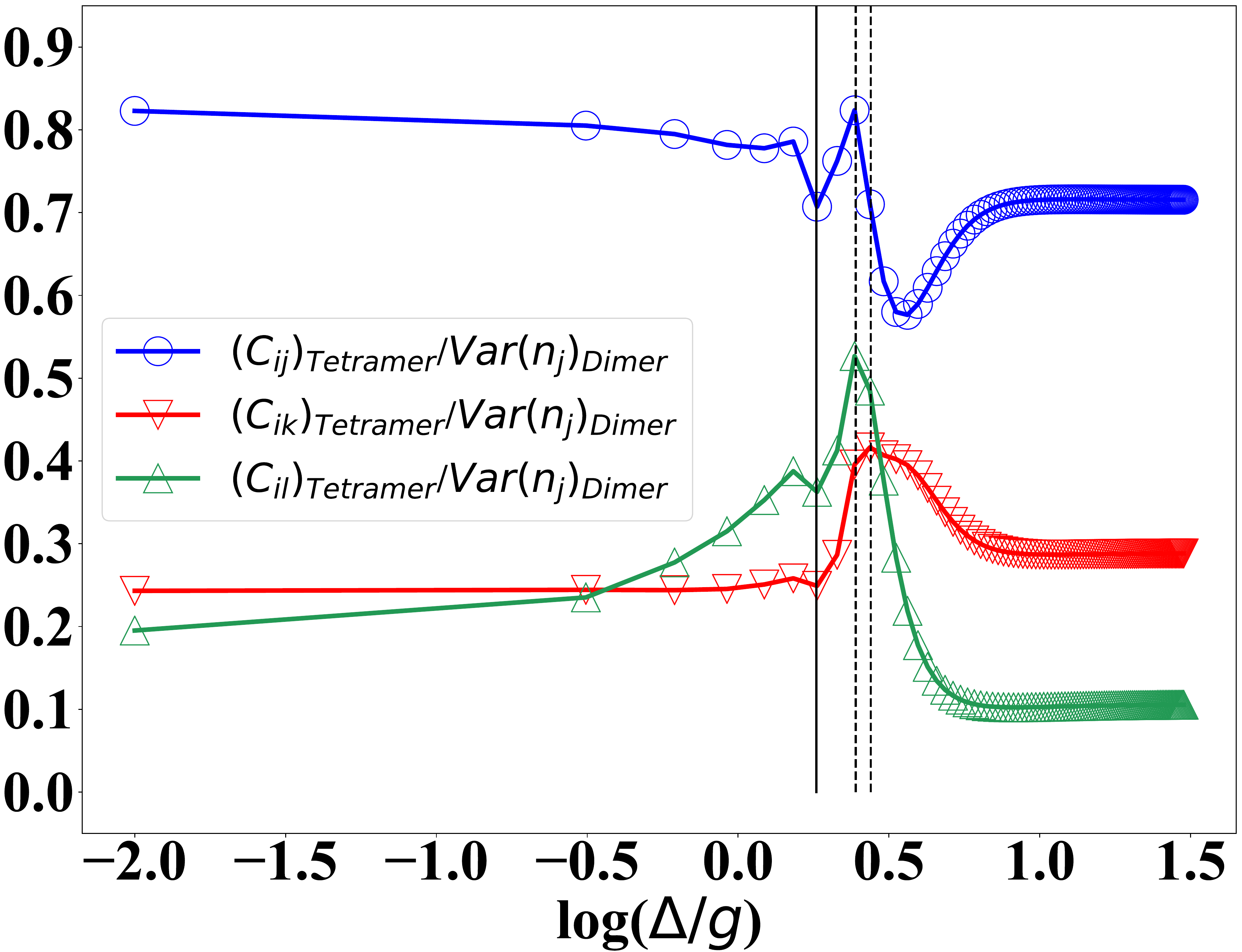}
\caption{Ratio between the absolute value of time-averaged two-point correlation functions $C_{ij}$, $C_{ik}$, $C_{il}$ of the tetramer (four-sites JC lattice), and the dimer variance in Eq.~(\ref{varianceI}). Vertical dashed lines, from left to right, indicate critical values of detuning $\Delta/g=(2.43, 2.73)$ where dynamical dimerization processes happen. Continuous vertical line stands for the anti-resonance. We use parameters $g=10^{-2}\omega$, $J=10^{-4}\omega$, where $\omega$ is the resonator frequency, and we consider up to $5$ Fock states per resonator.}
\label{Fig6}
\end{figure}

Let us discuss the results for the tetramer. Figure~\ref{Fig6} shows the ratio between the absolute value of time-averaged two-point correlation functions $C_{ij}$, $C_{ik}$, $C_{il}$ of the tetramer, and the dimer variance in Eq.~(\ref{varianceI}). We order the lattice sites from left to the right according to indexes $(i,j,k,l)$. It is noticeable that a larger JC lattice also exhibits resonances at critical values of detuning $\Delta/g=(2.43, 2.73)$ (vertical dashed lines), and the anti-resonance at $\Delta=1.82g$ (continuous vertical line). Moreover, the two-point correlation function $C_{il}$, associated with edges of the lattice, has the same resonance ($\Delta=2.43g$) as compared with nearest-neighbor correlation function $C_{ij}$. The latter suggests that for a finite one-dimensional JC lattice of $L$ sites, the number of resonances associated with dimerization processes corresponds to a universal number of different connectivities of the lattice, that is, connectivity $\nu=1$ for lattice edges, and connectivity $\nu=2$ for bulk lattice sites. These results are a consequence of the broken translational symmetry.

\subsection{Open system}
\label{open}

A realistic implementation of a strongly-correlated light-matter system should consider the system-bath interaction which leads to loss mechanisms in the initial state preparation and along the dynamics, {\it e.g.}, if we consider a experimental realization based on superconducting circuits Refs.~\cite{PhysRevX.4.031043,PhysRevX.7.011016}. In these experiments, the dissipative dynamics is described by a Markovian Lindblad master equation ($\hbar=1$)
\begin{align}
\frac{d\rho}{dt}=-i[H_{\rm{JCH}},\rho]+\sum^L_{i=1}\bigg(\gamma\mathbb{L}[\sigma^{-}_i]\rho+\gamma_\phi\mathbb{L}[\sigma^{z}_i]\rho+\kappa\mathbb{L}[a_i]\rho\bigg),
\label{Master}
\end{align} 
where the Liouvillian operator reads $\mathbb{L}[\mathcal{O}]\rho=\mathcal{O}\rho\mathcal{O}^{\dag}-\frac{1}{2}\{\mathcal{O}^{\dag}\mathcal{O},\rho\}$. We consider the same loss mechanisms for each lattice site including energy relaxation, dephasing; and photon losses at rates $\gamma$, $\gamma_{\phi}$, and $\kappa$, respectively. 

In order to prepare the initial state, we propose to include an ancillary two-level system on each lattice site, which interacts with the cavity mode. In this case, the Hamiltonian describing a single lattice site reads
\begin{align}
H_i = H^i_{\rm{JC}}+\omega_A\sigma^{+}_{A_i}\sigma^{-}_{A_i} + g_A(\sigma^{+}_{A_i}a_i+\sigma^{-}_{A_i}a^{\dag}_i),
\label{ancilla-site}
\end{align}
where $\omega_A$ is the ancilla frequency, $g_A$ the ancilla-cavity coupling strength, and $H^i_{\rm JC}$ the JC Hamiltonian of site $i$. 

The initialization protocol makes use of Gaussian and Stark pulses as described in Ref.~\cite{Majer:2007aa}. First, we let the system to cold down to its ground state $\ket{\psi_0}=\bigotimes^L_{i=1}\ket{0,-}_i\ket{\downarrow}_{A_i}$. Second, we apply individual Gaussian $\pi$ pulses acting upon each ancilla TLS in order to prepare the state $\bigotimes^L_{i=1}\ket{0,-}_i\ket{\uparrow}_{A_i}$. Third, a Stark pulse is applied to each ancilla TLS bringing it into resonance with its respective lattice site, {\it i.e.} $\omega_{A}=E^-_1$, during a time interval $\Delta\tau=\pi/(2 g_A t^{--}_1)$. In this way, the strong lattice site-ancilla interaction governed by Eq.~(\ref{ancilla-site}) yields the desired initial state $\ket{\psi_0}=\bigotimes^L_{i=1}\ket{1,-}_i\ket{\downarrow}_{A_i}$. Notice that one should satisfy the condition $|E_{1}^{+} - E_{1}^{-} | \gg g_{A}$. The latter avoids unwanted population of the state $\bigotimes^L_{i=1}\ket{1,+}_i\ket{\downarrow}_{A_i}$. Then, the ancilla-site interaction is suppress by applying a Stark pulse tuning $\omega_A$ below the frequency $E^-_1$. We stress that Stark pulses can be implemented by means of external magnetic fluxes applied upon superconducting quantum interference devices that form a transmon qubit~\cite{PhysRevA.76.042319,PhysRevB.77.180502}.    
\begin{figure}[t]
\centering
\includegraphics[scale=0.3]{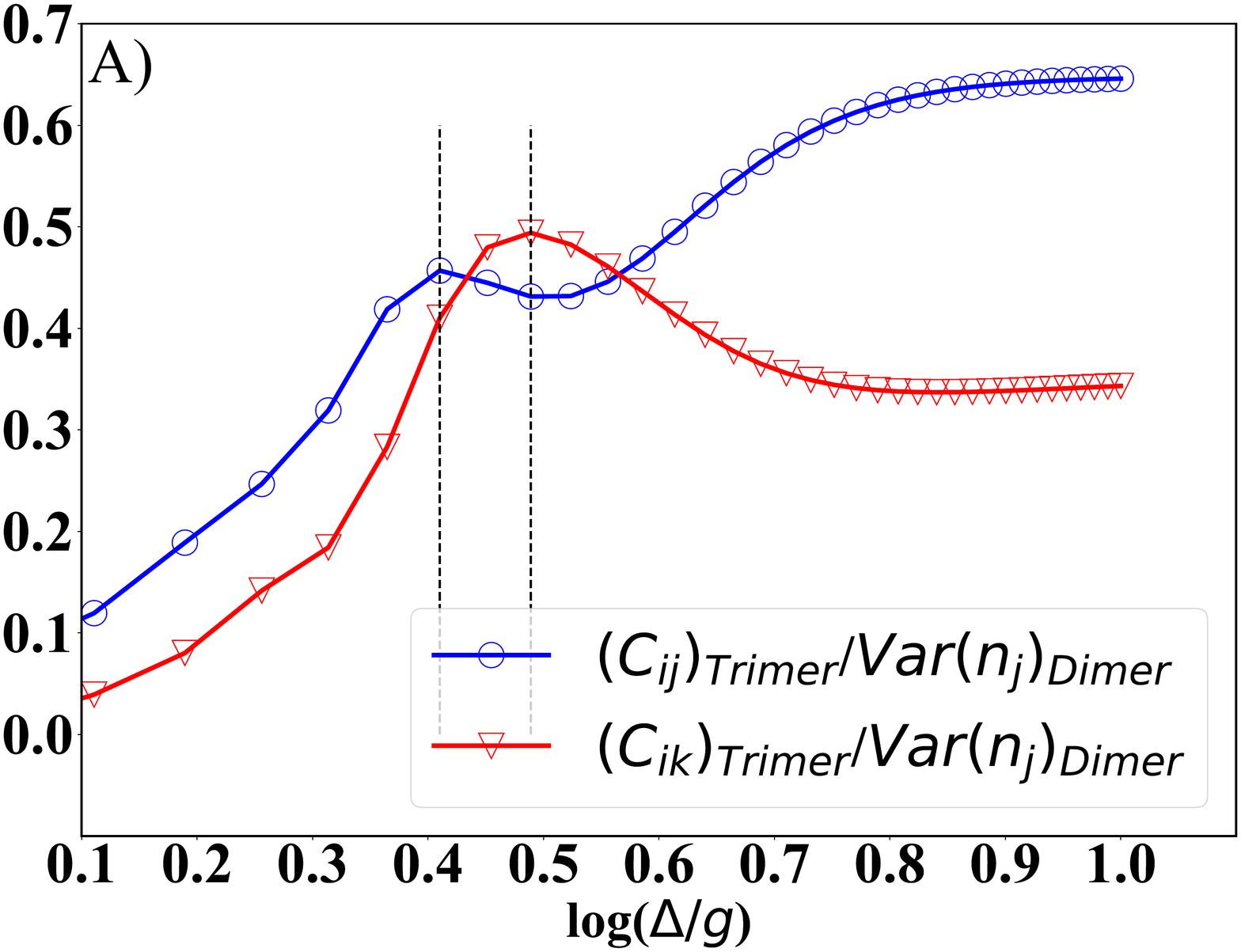}\\
\includegraphics[scale=0.3]{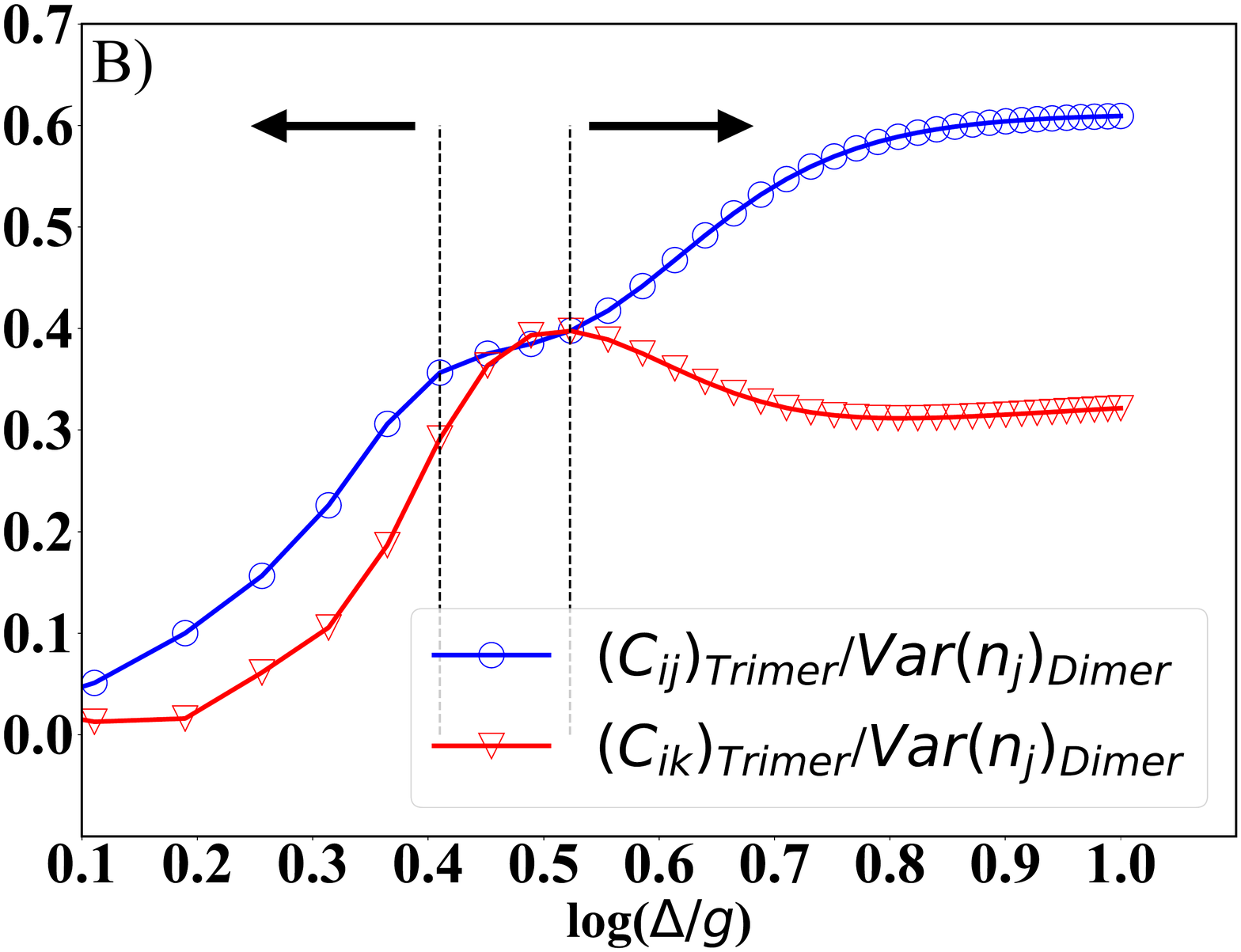}
\caption{Ratio between the absolute value of time-averaged nearest-neighbor correlation function $C_{ij}$ (next nearest-neighbor correlation function $C_{ik}$) of the trimer, and the dimer variance. (A) Vertical dashed lines, from left to right, indicate critical values of detuning $\Delta/g=(2.57, 3.08)$ where dynamical dimerization processes happen. Here, we use parameters $\nu_c=5$ GHz (cavity frequency), $g=200$ MHz, $J_f=2$ MHz, $\kappa=225$ KHz, $\gamma=35$ KHz ($T_1=28~\mu$s), and $\gamma_{\phi}=45$ KHz ($T_2=22~\mu$s). (B) Vertical dashed lines, from left to right, indicate critical values of detuning $\Delta/g=(2.57, 3.33)$. We use $\gamma=530$ KHz ($T_1=1.87~\mu$s), and $\gamma_{\phi}=450$ KHz ($T_2=2.22~\mu$s), other parameters remain the same. In these numerical calculations we consider up to $4$ Fock states per resonator.}
\label{Fig7}
\end{figure} 

We have numerically calculated the initial state preparation and the sudden quench using the Eq.~(\ref{Master}), using physical parameters taken from state-of-the-art circuit QED setups~\cite{PhysRevX.4.031043,PhysRevX.7.011016,10.1117/12.2192740,PhysRevB.77.180502}. Figure~\ref{Fig7}(a,b) shows the ratio $(C_{ij})_{\rm Trimer}/{\rm Var}(n_j)_{\rm Dimer}$ as a function of $\Delta/g$, where $(C_{ij})_{\rm Trimer}$ stands for the absolute value of the time-averaged correlation function, and ${\rm Var}(n_j)_{\rm Dimer}$ corresponds to the variance of the dimer. As seen in Fig.~\ref{Fig7}(a), we identify resonances corresponding to the critical values $\Delta/g=(2.57, 3.33)$, vertical dashed lines, within the intermediate regime of light-matter interaction, $1<\Delta/g<10$. In the same way, as in the closed system dynamics~\ref{closed}, the trimer experiences dynamical dimerization processes, where short- and long-range correlations dominate over dissipation. As we increase the energy relaxation and dephasing rates of TLSs, resonance peaks are spreading throughout to a wider detuning range, see Fig.~\ref{Fig7}(b). These results show evidence about the stability of dynamical dimerization processes that happen in a finite-sized Jaynes-Cummings lattice, and allow to establish a parameter threshold for the appearance of DDP in the dissipative case.  

In the numerical calculations we use $\nu_c=5$ GHz (cavity frequency), $g=200$ MHz, $J_f=2$ MHz, $\kappa=225$ KHz, $\gamma=35$ KHz ($T_1=28~\mu$s), and $\gamma_{\phi}=45$ KHz ($T_2=22~\mu$s)~\cite{10.1117/12.2192740} for Fig.~\ref{Fig7}(a), and $\gamma=530$ KHz ($T_1=1.87~\mu$s), and $\gamma_{\phi}=450$ KHz ($T_2=2.22~\mu$s)~\cite{PhysRevB.77.180502} for Fig.~\ref{Fig7}(b).

\section{Conclusions}
 \label{conclusion}
 In summary, we have reported on the emergence of a \emph{dynamical dimerization phase} in a finite-sized Jaynes-Cummings lattice as a result of a quantum quench from an initial state with integer filling factor. We have thoroughly analyzed the quench dynamics in a close two-sites Jaynes-Cummings lattice, which allows us to obtain analytical results for time-averaged order parameters such as the local variance of the number of polaritons, and the linear entropy. Further, these order parameters can be used to analyze and predict the resulting quenched dynamics for more complex architectures. When comparing the dimer variance with two-point correlation functions of the trimer and tetramer, it allows us to determine critical values for the detuning where dynamical dimerization processes happen. Recognizing resonances and anti-resonance for detuning values, in turn, allow controlling what kind of correlation dominates over the dynamics be short-range or a combination of both short- and long-range, and may also allow controlling polariton propagation along the lattice. We stress that the intrinsic dimer dynamics, characterized by the time-averaged variance (\ref{varianceI}), governs the quench dynamics of closed finite-sized JC lattices, and we expect similar results as one increases the number of lattice sites. 
 
In a realistic situation, it is necessary to include dissipative mechanisms in the state preparation and over the quench dynamics. Considering parameters of state-of-the-art circuit QED technology, permit numerical results to show that as one increases the coherence times of two-level systems and cavities, two sharp resonance peaks become more evident. These results demonstrate that DDP remains in the dissipative case. We conjecture that for a finite one-dimensional JC lattice of $L$ sites, the number of resonances associated with dimerization processes corresponds to the number of different connectivities of the lattice, that is, connectivity $\nu=1$ for lattice edges, and connectivity $\nu=2$ for bulk lattice sites. Our findings could be tested with state-of-the-art quantum technologies. For instance, in trapped ions technology, the Jaynes-Cummings-Hubbard model has been theoretically proposed in Ref.~\cite{PhysRevA.80.060301} and physically implemented in Ref~\cite{PhysRevLett.111.160501}. In superconducting circuits, the JC dimer has been implemented in Ref.~\cite{PhysRevX.4.031043}. In this case, homodyne signal detection may allow measuring the local variance of photon number, which can also be used as an order parameter.

\section*{Acknowledgements}
R. Pe\~na acknowledges the support from Vicerrector\'ia de Postgrado USACH, F. Torres acknowledges financial support from grants FA9550-16-1-0122, FA9550-18-1-0438, Fondecyt 1160639, and CEDENNA through the Financiamiento Basal para Centros
Cient\'icos y Tecnol\'ogicos de Excelencia-FB0807. G.Romero acknowledges the support from the Fondo Nacional de Desarrollo Cient\'ifico y Tecnol\'ogico (FONDECYT) under grant No. 1190727.

\bibliography{Mybib}

\end{document}